# Management von Sensordaten im Smarthome

## Besonderheiten und Ansätze

*Albrecht Kurze[1], Karola Köpferl[2], Andy Börner[1]*

1  Professur Medieninformatik, TU Chemnitz, Deutschland

2  Juniorprofessur Soziologie mit dem Schwerpunkt Technik, TU Chemnitz, Deutschland

{albrecht.kurze, karola.koepferl, andy.boerner}@informatik.tu-chemnitz.de

**Abstract**

Eine Vielzahl einfacher Sensoren, z. B. für Temperatur, Licht oder Luftfeuchte, zieht ins Smarthome ein. Dabei gibt es für Daten dieser Sensoren einige Besonderheiten zu beachten: a) aus der Natur der erfassten Messgrößen als "thin but big data", die einordnungs- und interpretationsbedürftig sind, b) wozu sowohl Algorithmen als auch Menschen fähig sind (wobei sich im Kontext des Zuhauses ein umfassender Informationswert ergibt, bis bin zur Erkennung von Aktivitäten, Verhalten und Gesundheit der Bewohner:innen), c) Nutzungen zu interessanten positiven Anwendungsfällen, aber auch zu missbräuchlichen Nutzungen und Einschränkungen der Privatsphäre führen. Beim Management solcher Daten ist die Berücksichtigung dieser Besonderheiten nötig, wofür Prinzipien der User Experience, Human-Data Interaction und des Datenschutzes gemeinsam betrachtet werden sollten. Wir stellen unser Forschungswerkzeug „Sensorkit" und den damit genutzten partizipativen Forschungsansatz zur Sammlung von Sensordaten in echten Wohnungen vor. Wir gehen in unseren Ergebnissen auf identifizierte Herausforderungen ein und erläutern, wie wir diese exemplarisch adressieren durch a) sinnvolle Voreinstellungen, b) Einsichts- und Eingriffsmöglichkeiten für Nutzende und c) Life-Cycle-Management der Daten. Dabei sind Phasen vor, während und nach der Sammlung, Verarbeitung und Nutzung der Sensordaten sowie die Bereitstellung nutzerfreundlicher Werkzeuge und Einbindung der Nutzenden wichtige Aspekte. Unsere Ergebnisse informieren über den Rahmen eines Forschungsprojekts und -werkzeugs hinaus auch die Entwicklung und Nutzung kommerzieller Smarthome-Geräte und -Services.

# 1    Einleitung und Besonderheiten

Die Anzahl an Smarthome-Geräten steigt kontinuierlich (acatech 2023). Besonders immer kleinere und günstigere einfache Sensoren werden zunehmend eingesetzt, auch in Haushaltsgeräten oder integriert in die Wohnumgebung selbst. Für Daten und Kontext dieser Sensoren sind einige Besonderheiten zu beachten.

Sensordaten sind als „thin but big data" (Gomez et al. 2022) einzuordnen: Sie umfassen große Mengen, doch ein einzelner Datenpunkt hat wenig Informationswert und ist interpretationsbedürftig. Ihre Bedeutung entsteht aus dem Kontext, etwa durch Zeitreihen oder den Vergleich mit anderen Messgrößen. Sensoren in Smarthome-Geräten, Haushaltsgeräten oder der Wohnumgebung sammeln Daten über Jahre oder Jahrzehnte, was potenziell Milliarden von Datenpunkten generiert, die aber selbst mit üblicher Technik problemlos speicherbar wären. Menschen und Algorithmen können diese Daten interpretieren (Laput et al. 2017; Tolmie et al. 2016), wobei Menschen auf situiertes Wissen zur Wohn- und Sozialsituation sowie Alltagswissen zurückgreifen, um den Kontext zu verstehen. Schon wenig Kontextwissen ermöglicht plausible Interpretationen und Rückschlüsse auf Anwesenheit, Aktivitäten oder Gesundheit der Bewohner:innen (Kurze et al. 2020).

Smarthome-Sensordaten bieten Anwendungsmöglichkeiten für Komfort, Sicherheit und Effizienz (Quack/Liu/Gröger 2019), bergen aber auch Risiken wie Bevormundung und Überwachung (Berger et al. 2023) – auch lateral zwischen Bewohner:innen. Transparenz und Kontrolle über die eigenen Daten sind hierbei zentrale Herausforderungen (Kurze et al., 2020). Die Wohnung als sozialer Kontext spielt eine besondere Rolle, da die Daten „situiert" sind und unter Berücksichtigung der sozialen und moralischen Ordnung interpretiert werden (Suchman 2006). Nutzende unter- oder überschätzen dabei oft die Aussagekraft, da plausible Erklärungen als intersubjektive Wahrheit akzeptiert werden, unabhängig von objektiver Wahrheit (Kurze et al. 2020).

Obwohl nur „einfache" Sensordaten, stellen sie „bösartige" Probleme dar, die keine trivialen Lösungen zulassen (Kurze/Bischof 2021). Für handelsübliche Smarthome-Geräte ist es meist notwendig, dass Nutzende sich registrieren und akzeptieren, dass Daten in eine Hersteller-Cloud geladen werden. Dabei entstehen Probleme oft auch unabhängig davon, wo und wie genau die Daten gespeichert oder verarbeitet werden, z. B. schon bei rein innerhäuslicher Nutzung ohne KI-Auswertung und ohne Dritte wie Tech-Giganten oder

Vermieter-Cloud (Börner et al. 2024). Diese Faktoren verstärken die Probleme potenziell weiter, z.B. durch Zusammenführung mit weiteren Daten wie Personen-, Account- und Nutzungsdaten (Friedewald/Karaboga/Zoche 2015).

Dabei bleibt oft unklar: Welche Daten werden erhoben? Mit welchen weiteren Daten und Metadaten werden sie verknüpft und wo werden sie gespeichert? Wie kann man diese Daten verwalten?

Es ist offen, wie Smarthome-Sensordaten juristisch eingeschätzt werden (Johns 2019), ob als personenbezogene Daten oder zumindest als personenbeziehbar, da sie prinzipiell (mit Zusatzwissen oder Verknüpfung) o.g. Rückschlüsse ermöglichen und damit einen Personenbezug aufweisen. Entsprechend umsichtig sollte auch unter diesem Aspekt mit den Daten verfahren werden.

Ein Management dieser Daten muss daher die Besonderheiten beachten und sollte nicht nur im Hinblick auf technische Machbarkeit und Handhabbarkeit erfolgen, sondern auch etablierte Prinzipien und Ansätze zu

- Usability/UX wie Aufgabenangemessenheit und Steuerbarkeit (ISO 9241-110),
- Human-Data Interaction wie Legibility, Agency und Negotiability (Mortier et al. 2015) und
- Datenschutz wie Datensparsamkeit, Zweckbindung und Privacy by Design (Art. 5 DSGVO)

berücksichtigen.

Bisherige Arbeiten adressieren die Besonderheiten entweder nicht zentral oder bleiben theoretisch, z.B. Kilic et al. (2021) in einem Ideation-Kartenspiel, aber ohne echte Sensordaten, zu Technologien zum Self-Management persönlicher Daten (Marillonnet et al. 2021), aber ohne Fokus auf die Wohnung, oder in DataBox (Mortier et al. 2016) ohne aktive Weiterentwicklung.

## 2  Forschungsansatz und -werkzeug Sensorkit

Unserer Forschung nutzt unser selbstentwickeltes „Sensorkit" zur Erfassung, Verarbeitung und Visualisierung von Sensordaten. Das Sensorkit umfasst Sensoren für Licht, Temperatur, Luftfeuchtigkeit, Bewegung, Luftqualität, $CO_2$ und Lautstärke, die Daten drahtlos an einen Raspberry Pi senden (Kurze 2022). Die Daten werden rein lokal verarbeitet (NodeRed), in einer Zeitrei-hendatenbank (InfluxDB) gespeichert und können auf einem Tablet visualisiert werden (Grafana). Mit den gewählten Einstellungen werden ca. 350.000 Datenpunkte (ca. 3 MB) pro Tag generiert, was ca. 100 Mio. Datenpunkte / 1 GB Daten pro Jahr bedeuten würde – eine heutzutage auch für Jahrzehnte leicht speicherbare Datenmenge.

In Feldstudien werden die Sensorkits für ca. zehn bis 14 Tage in Wohnungen installiert, um Nutzende mit den Daten interagieren zu lassen. Im Anschluss werden Workshops durchgeführt. Dieser partizipative Ansatz ermöglicht Einblicke in die individuelle und kollektive Datenarbeit und hilft, mögliche Risiken zu identifizieren (Kurze et al. 2020) und auf mögliche Probleme bei der Benutzung smarter Technik hinzuweisen (Köpferl et al. 2024).

## 3  Ansätze und Beispiele im Sensorkit

Im Sensorkit haben wir verschiedene Ansätze zum Management von Smarthome-Sensordaten exemplarisch umgesetzt. Dabei sind diese Punkte nicht erschöpfend, sondern Gegenstand fortlaufender Forschung und Entwicklung.

**Sinnvolle Voreinstellungen durch Hersteller zu/von**
- Messgrößen und Parameter in der Erfassung der Sensordaten
- neutraler Zuordnung von technisch-organisatorischen Metadaten, um die Privatsphäre zu schützen
- Verarbeitung der Sensordaten (genutzte Metriken und Auswertungen)
- Speicherung der Sensordaten (z. B. automatisierte Aggregationen und Löschfristen, um unerwünschte Nachnutzung zu verhindern).

**Transparenz-, Kontroll- und Interaktionsmöglichkeiten für Nutzende zu**
- Einstellungen zu den vorherigen Punkten
- vollständiger Einsicht in erfasste Daten, in sinnvoller Kontextualisierung, z. B. in Zeitreihen und von Daten untereinander
- Datenhaltung und -löschung (einfach/vollständig und selektiv/feingranular)
- Zugriffsrechten, diese sollten klar definiert und beschränkt werden, z. B. mit Role Based Access Control und Maßnahmen für gemeinsamen Zugriff, z. B. Vier-Augen-Prinzip im Haushalt
- Datenweitergabe und -teilen, insbesondere zu Dritten mit genau definierten Zielen und Zwecken

- Unterstützung der eigenen Sinngebung und Reflexion zu Implikationen, z.B. mit eigenen Metadaten und Annotationen sowie durch kollektive Datenauseinandersetzung zum Aufbau von Data Literacy.

**Life-Cycle-Management** (für Jahre und Jahrzehnte der Nutzung mit Berücksichtung von Phasen vor, während und nach der eigentlichen Nutzung)

- Datenportabilität und Weiternutzung (Backup, Export, Import, z.B. bei Defekt oder Wechsel)
- Weitergabe/Austausch von Geräten/System (Verkauf, Neubezug)
- Nachnutzung und Vererbung von Systemen und Daten (digitaler Nachlass und Re-Initialisierung).

## 4 Zusammenfassung und Ausblick

Ein gutes Management von Smarthome-Sensordaten unterstützt nicht nur positive Nutzungsziele wie Effizienz und Sicherheit, sondern stärkt auch Werte wie Autonomie, Kontrolle, Transparenz und Privatheit. Die Einbindung der Nutzenden und sinnvolle Werkzeuge sind dabei wichtige Aspekte. Sinnvolles Datenmanagement adressiert daher Entwickler*innen und Nutzende gleichermaßen. Wie exemplarisch am Sensorkit gezeigt, können Forschungsansätze und -werkzeuge die Nutzung echter Smarthome-Technik informieren.

**Acknowledgements**

Diese Arbeit ist gefördert vom Bundesministerium für Bildung und Forschung (BMBF) im Rahmen des Projektes *Simplications*, FKZ 16KIS1868K.

## Literatur